\documentclass[12pt]{iopart}
\usepackage{graphicx}
\usepackage{amssymb}

\begin{document}
\title{Backward energy flow in simple 4-wave electromagnetic fields}
\author{Peeter Saari$^{1,2}$ and Ioannis Besieris$^{3}$}
\address{$^{1}$Institute of Physics, University of Tartu, W. Ostwaldi 1, 50411, Tartu, Estonia}
\address{$^{2}$Estonian Academy of Sciences, Kohtu 6, 10130 Tallinn, Estonia}
\address{$^{3}$The Bradley Department of Electrical and Computer Engineering, Virginia
Polytechnic Institute and State University, Blacksburg, Virginia 24060, USA}
\date{\today}

\vspace{10pt}
\ead{Corresponding author peeter.saari@ut.ee}
\vspace{10pt}
%\begin{indented}
%\item[] February 2021
%\end{indented}

\begin{abstract}
Electromagnetic energy backflow is a phenomenon occurring in regions where the direction of the Poynting vector is opposite to that of the propagation of the wave field.  It is particularly remarkable in the nonparaxial regime and has been exhibited in the focal region of sharply focused beams, for vector Bessel beams, and vector-valued spatiotemporally localized waves. A detailed study is undertaken of this phenomenon and the conditions for its appearance are examined in detail in the case of a superposition of four plane waves in free space, the simplest electromagnetic arrangement for the observation of negative energy flow, as well as its comprehensive and transparent physical interpretation. It is shown that the state of polarization of the constituent components of the electromagnetic plane wave quartet determines whether energy backflow takes place or not and what values the energy flow velocity assumes. Depending on the polarization angles, the latter can assume any value from $c$ (the speed of light in vacuum) to $-c$ in certain spatiotemporal regions of the field.
\end{abstract}

\vspace{10pt}
\begin{indented}
\item[]Keywords: Poynting vector, energy velocity, polarization, energy backflow
\end{indented}

%\maketitle   See teeb tiitellehe

\section{Introduction}

Energy backflow of a free-space electromagnetic (EM) field is an unusual effect 
exhibited in regions where the direction of the Poynting vector is opposite to 
the propagation direction of the wave field. 
In this sense, the terms 'reversed' or 'negative' Poynting
vector are also used. The effect is particularly remarkable in the case of
non-paraxial propagation and occurs, e.g., in the focal region of sharply
focused optical beams \cite{negSfocal}-\cite{negSfocaluus2}, in
superpositions of four plane waves directed under equal angles $\theta$ with
respect to the propagation axis \cite{Katz,quartet}, for vector Bessel
beams \cite{AriBBnegS}-\cite{BBpull} and vector X-waves \cite{TM+TExwave}. 
Note that Bessel beams are superpositions of infinitely many monochromatic plane 
waves making an angle $\theta$ with respect to a specific axis (say $z$) which 
corresponds to the direction of propagation of the Bessel beam, and 
X-waves---representatives of the rich family of so-called localized waves---are 
just the same superpositions of ultrashort pulses \cite{LWs}-\cite{AbourNature}. 
Energy backflow has attracted much interest in
connection with applications in microparticle manipulation like 'tractor
beams', etc., see reviews \cite{pullReview,pullRevUus}.

Our purpose is to study in detail the mechanism and conditions for the appearance
of the reversed Poynting vector in a quartet of plane waves---the simplest of
EM fields for observation of the effect. In distinction from earlier studies,
we consider time-dependent EM fields instead of common cycle-averaged ones, and
make use of the energy flow velocity. Although the latter is given by a
well-known formula related to the Poynting vector and the energy density, it 
exhibits some surprising properties. We hope that such a study will result in a 
more comprehensible and transparent picture of the nature of the backward energy 
flow effect than it can be deduced from cumbersome formulas describing the effect 
in the case of the vector Bessel beams and other more complicated fields.

In the next two sections we shall study a quartet of plane waves and show how
the polarizations of the constituents of the quartet determine whether the
energy backflow takes place or not, and what values the energy flow velocity
assumes. Section IV is devoted to studying of a general dependence of the energy 
flow velocity on the polarization angles.
In section V we discuss the results, touch on the problem
of physical interpretation of the Poynting vector and some alternative
definitions---noteworthy in the given context---of the energy
flow velocity. Concluding remarks are provided in section VI.

\section{4-wave fields with regions of negative energy velocity}

A quartet of plane waves symmetrically directed with respect to the
propagation axis of the whole field constitute the simplest EM field for which 
energy backflow is possible\footnote{There is a claim in the literature
\cite{pullReview} that a superposition of two plane waves is sufficient to
introduce a negative Poynting vector. This is not wrong, but in this case
the vector is not reversed with respect to the propagation direction of the
resultant field.}. In order to study the superposition of four interfering plane
waves of different polarizations, we start from a "seed" plane wave propagating
along the axis $z$ with the wavevector $\mathbf{k}_{0}=\left(  0,0,k\right)  $
and polarization vector $\mathbf{e}_{0}\left(  \phi\right)  =\left(  \cos
\phi,\sin\phi,0\right)  $, where $\phi$ is an arbitrary angle of (linear)
polarization with respect to the axis $x$. By making use of well-known
rotation $\left(  3\times3\right)  $-matrices $\mathbf{R}_{x}\left(
\theta\right)  $ and $\mathbf{R}_{y}\left(  \theta\right)  $ which rotate
vectors by an angle $\theta$ about the $x$- and $y$-axis, respectively, we
express the wave vectors of our quartet of plane waves (see figure 1) as
follows:%
\begin{eqnarray}
\mathbf{k}_{1}  & =\mathbf{R}_{y}\left(  \theta\right)  \mathbf{k}%
_{0}\;,\nonumber\\
\mathbf{k}_{2}  & =\mathbf{R}_{x}\left(  -\theta\right)  \mathbf{k}%
_{0}\;,\label{4k}\\
\mathbf{k}_{3}  & =\mathbf{R}_{y}\left(  -\theta\right)  \mathbf{k}%
_{0}\;,\nonumber\\
\mathbf{k}_{4}  & =\mathbf{R}_{x}\left(  \theta\right)  \mathbf{k}%
_{0}\;.\nonumber
\end{eqnarray}

In this section we study the case where the "seed" wave for all four waves is
polarized along the axis $x$, i.e., $\phi=0$. Consequently, we obtain the
following polarization vectors in the quartet:%
\begin{eqnarray}
\mathbf{e}_{1}  & =\mathbf{R}_{y}\left(  \theta\right)  \mathbf{e}_{0}\left(
0\right)  =\left(  \cos\theta,0,-\sin\theta\right)  ,\nonumber\\
\mathbf{e}_{2}  & =\mathbf{R}_{x}\left(  -\theta\right)  \mathbf{e}_{0}\left(
0\right)  =\left(  1,0,0\right)  ,\label{4e}\\
\mathbf{e}_{3}  & =\mathbf{R}_{y}\left(  -\theta\right)  \mathbf{e}_{0}\left(
0\right)  =\left(  \cos\theta,0,\sin\theta\right)  ,\nonumber\\
\mathbf{e}_{4}  & =\mathbf{R}_{x}\left(  \theta\right)  \mathbf{e}_{0}\left(
0\right)  =\left(  1,0,0\right)  ,\nonumber
\end{eqnarray}
which are also depicted in figure 1.%

\begin{figure}[ptb]%
\centering
\includegraphics[width=8cm]{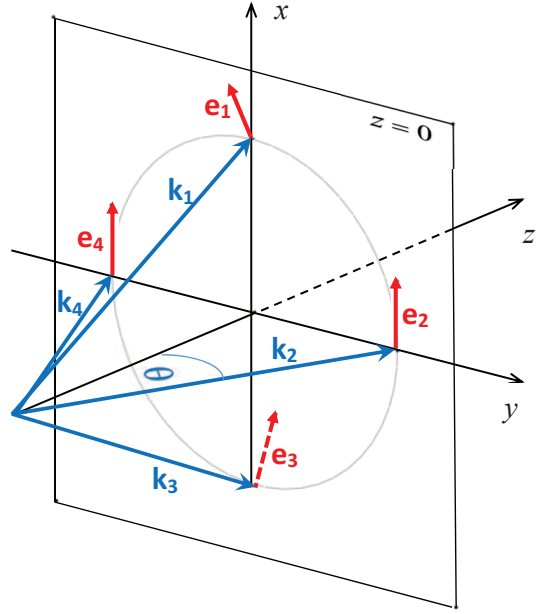}
\caption{Wave and polarization vectors for the quartet of plane
waves according to Eqs.~(\ref{4k}) and Eq.~(\ref{4e}). }%
\label{Joon1skeem}%
\end{figure}

Our aim is to study not only spatial but also temporal dependence of energy
flow, i.e. we cannot restrict ourselves to calculation of a cycle-averaged
Poynting vector. Therefore, instead of complex field expressions we have to
work with real ones. We assume a spatio-temporal dependence of the
constituent plane waves of the form $\cos\left(  \mathbf{kr}-kct\right)  $,
where $k=\omega /c=2\pi /\lambda$ 
and $\mathbf{r}=\left(  x,y,z\right)  $. As a matter of fact, if the cosine
is multiplied by a pulse envelope or replaced by any other suitable real
function describing the pulse, some final results will still hold. So we get
for the electric field of the quartet%

\begin{eqnarray}
\mathbf{E}\left(x,y,z,ct\right)   & =%
{\displaystyle\sum\limits_{j=1}^{4}}
\cos\left(  \mathbf{k}_{j}\mathbf{r}-kct\right)  \mathbf{e}_{j}=\nonumber\\
& \fl=\left(
\begin{array}
[c]{c}%
2\cos(kz\cos\theta-kct)\left[  \cos\left(  ky\sin\theta\right)  +\cos\left(
kx\sin\theta\right)  \cos\theta\right] \\
0\\
2\sin(kz\cos\theta-kct)\sin\left(  kx\sin\theta\right)  \sin(\theta)
\end{array}
\right), \label{Enurgad0}%
\end{eqnarray}
where unity amplitude of each plane wave has been assumed. Eq.~(\ref{Enurgad0})
indicates that the field propagates along the axis $z$ due to the
symmetry of the pairs of plane waves in the quartet. 
Likewise for the magnetic field we get%
\begin{eqnarray}
\mathbf{B}\left( x,y,z,ct\right)   
& ={\displaystyle\sum\limits_{j=1}^{4}} 
\cos \left( \mathbf{k}_{j}\mathbf{r}-kct\right) 
\left( \mathbf{k}_{j}\times \mathbf{e}%
_{j}\right) =  \nonumber \\
& \fl=\left(
\begin{array} [c]{c}%
0 \\ 
2\cos (kz\cos \theta -kct)\left[ \cos \left( kx\sin \theta \right) 
+\cos \left(
ky\sin \theta \right) \cos \theta \right]  \\ 
2\sin (kz\cos \theta -kct)\sin \left( ky\sin \theta \right) \sin (\theta )%
\end{array}%
\right), \label{Bnurgad0}
\end{eqnarray}%
where Heaviside--Lorentz units ($\varepsilon _{0}=\mu_{0}=1$) 
and normalized speed of light $c=1$ are assumed. 
In the given units
the Poynting vector $\mathbf{S}$ and the energy density $w$ are given by%
\begin{eqnarray}
\mathbf{S}\left( x,y,z,ct\right)  &=&\mathbf{E}\left( x,y,z,ct\right) \times 
\mathbf{B}\left( x,y,z,ct\right) ,  \nonumber \\
w\left( x,y,z,ct\right)  &=&\frac{1}{2}\left[ \mathbf{E}^{2}\left(
x,y,z,ct\right) + \mathbf{B}^{2}\left( x,y,z,ct\right) \right].
\label{Sw}
\end{eqnarray}%

For the quartet the expression of $\mathbf{S}$ turns out to be not too
cumbersome if one introduces the following rescaled and dimensionless
coordinates: $\widetilde{x}\equiv k~x\sin\theta$, $\widetilde{y}\equiv
k~y\sin\theta$, $\widetilde{z}\equiv k~z\cos\theta$, $\widetilde{t}\equiv kct$.
With these definitions
we obtain%
\begin{equation}
\mathbf{S}\left(  \widetilde{x},\widetilde{y},\widetilde{z},\widetilde{t}%
\right)  =\left(
\begin{array}
[c]{c}%
-4\cos\left(  \widetilde{z}-\widetilde{t}\right)  \sin\left(  \widetilde{z}%
-\widetilde{t}\right)  \sin\theta\sin\widetilde{x}~\left(  \cos\widetilde{x}%
+\cos\widetilde{y}\cos\theta\right)  \\
-4\cos\left(  \widetilde{z}-\widetilde{t}\right)  \sin\left(  \widetilde{z}%
-\widetilde{t}\right)  \sin\theta\sin\widetilde{y}~\left(  \cos\widetilde{y}%
+\cos\widetilde{x}\cos\theta\right)  \\
4\cos^{2}\left(  \widetilde{z}-\widetilde{t}\right)  \left(  \cos
\widetilde{x}+\cos\widetilde{y}\cos\theta\right)  \left(  \cos\widetilde{y}%
+\cos\widetilde{x}\cos\theta\right)
\end{array}
\right) .\label{Snurgad0}%
\end{equation}
From Eq.~(\ref{Snurgad0}) the following conclusions can be drawn:

\begin{enumerate}
\item the field of the Poynting vector propagates as a whole with velocity
$v=c/\cos\theta$ along the axis $z$ and it is invariant because it depends on time only through the 
difference $z\cos\theta-ct$ (we name it as the propagation variable);

\item if $z=t=0$ and generally on planes $k(z\cos\theta-ct)=n\pi
,~n\in\mathbb{Z}$ the transverse components of $\mathbf{S}$ vanish;

\item on planes $k(z\cos\theta-ct)=(2n+1)\pi/2,~n\in\mathbb{Z}$ the field
$\mathbf{S}$ vanishes;

\item the longitudinal component $S_{z}$ is invariant with respect to
interchange of coordinates $x$ and $y$;

\item the transverse components transform to each other if the coordinates $x$
and $y$ are interchanged;

\item there are regions in any transverse plane where the longitudinal
component $S_{z}$ is negative, i.e., the energy flows backward and this
phenomenon takes place irrespectively of time instant or cycle averaging,
although the latter affects the numerical value of negative $S_{z}$.
\end{enumerate}

The energy density is given as follows in terms of the dimensionless coordinates:

\begin{eqnarray}
w\left( \widetilde{x},\widetilde{y},\widetilde{z},\widetilde{t}\right) 
&=&2\cos ^{2}\left( \widetilde{z}-\widetilde{t}\right) [\left( 1+\cos
^{2}\theta \right) \cos ^{2}\widetilde{x}+4\cos \theta \cos \widetilde{x}%
\cos \widetilde{y}+  \nonumber \\
&&\left( 1+\cos ^{2}\theta \right) \cos ^{2}\widetilde{y}]+2\sin ^{2}\left( 
\widetilde{z}-\widetilde{t}\right) \left( \sin ^{2}\widetilde{x}+\sin ^{2}%
\widetilde{y}\right) \sin ^{2}\theta . \nonumber
%\label{wnegaS}
\end{eqnarray}

Similarly to $S_{z}$, the energy density obeys the symmetry $w\left(
x,y,z,ct\right)  =w\left(  y,x,z,ct\right)  .$ This conclusion is confirmed
by numerically calculated plots, an example of which is presented in figure 2.%

\begin{figure}[h]%
\centering
\includegraphics[width=16cm] {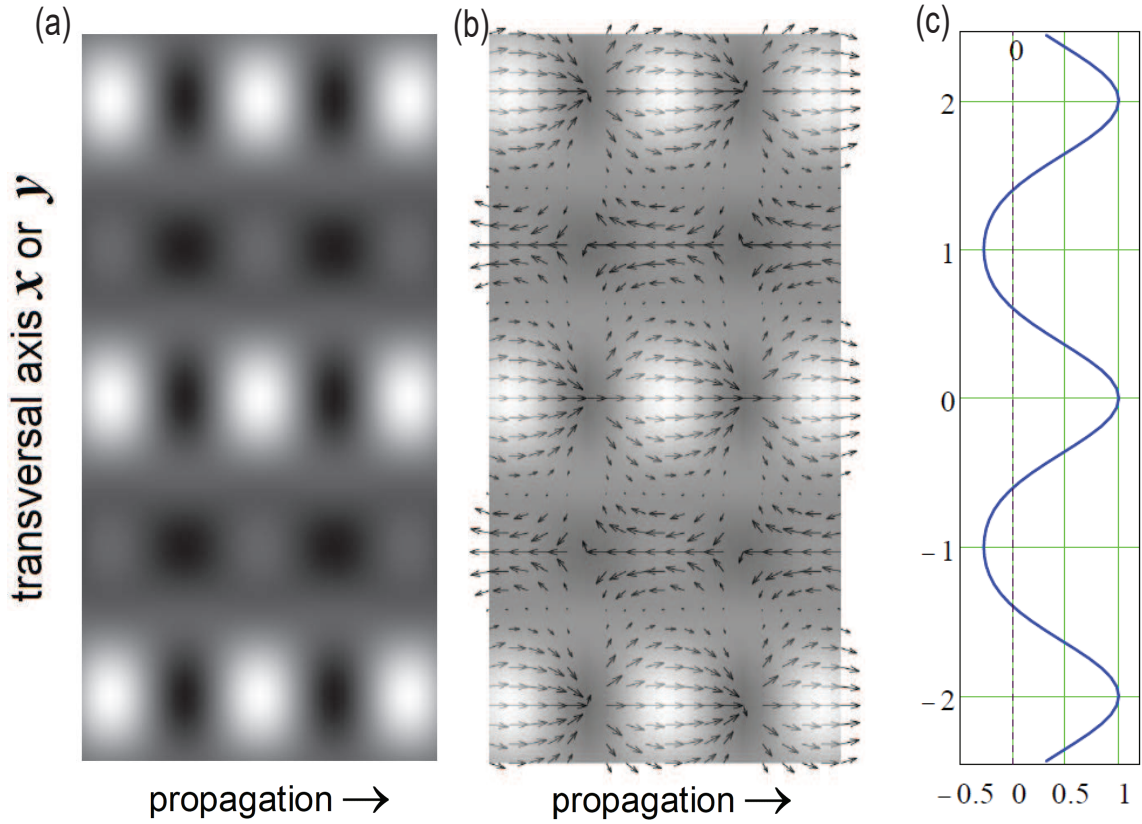}
\caption{Energy density and energy flow for the quartet
of plane EM waves with polarization vectors given by Eq.~(\ref{4e}) and
$\theta=2\pi/5=72^{\circ}$. (a)---greyscale plot of $w$ as a function of the
transverse coordinate $\protect\widetilde{x}$ at $\protect\widetilde{y}=0$
(or $\protect\widetilde{y}$ at $\protect\widetilde{x}=0$) and the propagation
variable $\protect\widetilde{z}-\protect\widetilde{t}$; (b)---the same with
decreased contrast and superimposed by a vector field plot of the energy flow
velocity. Shown are projections of the velocity vectors onto the plane
$\left(  z,x\right)  $; (c)---dependence on $\protect\widetilde{x}$ (along
vertical axis) of normalized longitudinal ($z$-) component of the Poynting
vector $S_{z}(\protect\widetilde{x},0,0,0)/S_{z}(0,0,0,0)$ showing the minimal
values ~$-0.28$. The range of the vertical axes is $[-2.5\pi,2.5\pi]$ and of the 
propagation axes is $[-1.25\pi,1.25\pi]$.
In plot (c) the scale of the transversal coordinate
$\protect\widetilde{x}=kx\sin\theta$ along the vertical is shown in units of $\pi$.}%
\label{Fig2}%
\end{figure}

The interference pattern in the energy density (figure 2a) remarkably differs from
that of two plane waves: if the pair (waves \# 2,4) were removed from the
quartet, all maxima would become of equal intensity (e.g., five equally bright
maxima along the central vertical $\widetilde{z}-\widetilde{t}=0$ instead of
three in figure 2a). Since the magnitude of the Poynting vector $\mathbf{S}$
is strong enough only in regions of local maxima and therefore a vector field plot of
$\mathbf{S}$ would be of low distinctness, we present the vector field plot
of the energy flow velocity $\mathbf{V}=\mathbf{S}/w$ instead. Figure 2b clearly
reveals that at locations of weak maxima of $\ w$ the energy flows
backward. While the strength of the component $S_{z}$ at these
locations comprises only 28 \% (see figure 2c) of that in locations of  strong 
maxima of $w$, the velocities $\mathbf{V}$ have exactly the same absolute value 1 (or
$c$ with dimensions) but the \textit{opposite signs}. Numerical evaluation
shows---and it is evident from figure 2b also---that wherever the velocity is
purely longitudinal it equals to the physically maximal possible value $\pm c$
(small irregularities in the vector field at points of vanishing $w$ are
computational artifacts due to "zero divide by zero"). We point out the very
noteworthy result that energy backflow occurs also with velocity $c$. While
according to Eqs.~(\ref{Enurgad0}) and (\ref{Bnurgad0}) neither of the fields
$\mathbf{E}$ or $\mathbf{B}$ is transverse, at locations where $V_{z}%
\mathbf{=}\pm c$ the fields are \textit{locally} TEM, mutually perpendicular
and of equal magnitude (in the chosen units)---as they have to be for a
\textit{null }EM field.

Note that the velocity components $V_{z}$ and $V_{x}$ on the plane $\left(
y=0\right)  $ constitute the vector field plot in figure 2b. If instead we
plotted figure 2b with $V_{z}$ and $V_{y}$ in the same plane, we would obtain
another projection of the energy flow velocity, where all arrows would be
directed with respect to the axis $z$ either parallel (in the regions of
stronger maxima of $w$) or antiparallel (in the regions of weaker maxima of
$w$). This is understandable because according to Eq.~(\ref{Snurgad0}) $S_{y}$
vanishes on the plane $\left(  y=0\right)  $.

Naturally, the smaller the angle $\theta$ the less remarkable will be the backflow
effect: peak negative values of the normalized $S_{z}$ decrease to $11\%$ if
$\theta=\pi/3$ and to $3\%$ if $\theta=\pi/4$. Of course, in the case of
paraxial geometry the energy backflow vanishes. However, the abovementioned
characteristic features of the velocity vector field remain unchanged with
decreasing $\theta$, except transverse narrowing of the backflow channels.
\begin{figure}[h]
\centering
\includegraphics[width=12cm] {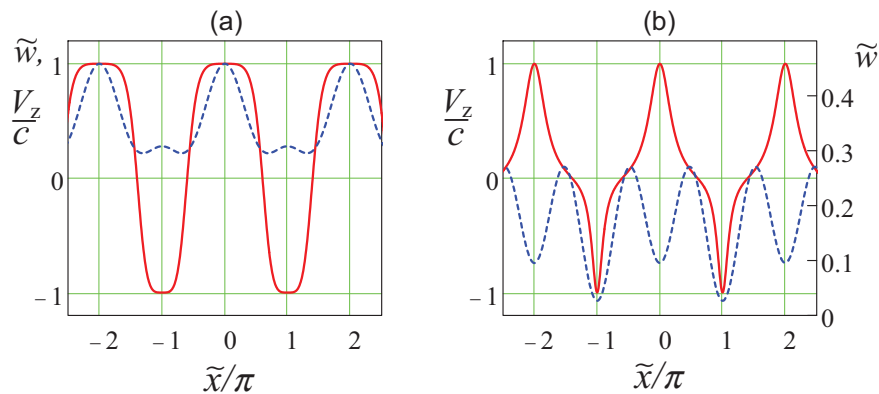} 
\caption{Dependence of the longitudinal component of the energy velocity $%
V_{z}$ (solid curve) and the normalized energy density (dashed curve) $%
\protect\widetilde{w}\equiv w(\protect\widetilde{x},\protect\widetilde{y},%
\protect\widetilde{z},\protect\widetilde{t})/w(0,0,0,0)$ on the
dimensionless transverse coordinate $\protect\widetilde{x}\equiv kx\sin 
\protect\theta $; (a)---along the line $y=0$ on the plane $(\protect%
\widetilde{z}-\protect\widetilde{t}=0)$ and (b)---along the line $y=0$ on
the plane $(\protect\widetilde{z}-\protect\widetilde{t}=0.4\protect\pi )$.
Note that in the plot (b) the scale for $\protect\widetilde{w}$ \ is on the
right-hand side and differs from the joint scale for $V_{z}$ and $\protect%
\widetilde{w}$ in (a).}
\end{figure}

Similar narrowing of the backflow channels is observable also if one moves
away from the planes $\widetilde{z}-\widetilde{t}=n\pi ,~n\in \mathbb{Z}$.
Curves in figure 3 show the dependence of the longitudinal component of the
energy flow velocity in comparison with that of the energy density along the
central vertical in figure 2b, i.e., where $\widetilde{z}-\widetilde{t}=0$.
We see in figure 3a that transverse widths of the regions where $V_{z}=\pm c$
are approximately equal to FWHM of the maxima of $w$. But on the plane $%
\widetilde{z}-\widetilde{t}=0.4\pi $, which is close to the plane $%
\widetilde{z}-\widetilde{t}=\pi /2$ where $V=0$, the backflow regions narrow
substantially and their FWHM become about two times less than those of the
minima or maxima of the energy density. By introducing the separate secondary 
scale for $\widetilde{w}\left(x,0,0.4\pi,0\right)=w\left(x,0,0.4\pi,0\right)
/w\left(0,0,0,0\right)$ on the right border of the plot, it is clearly seen 
in figure 3b that the negative minima of $V_{z}$ are encompassed by the (positive) 
minima of $\widetilde{w}$.

In the discussion so far it has been assumed that the seed polarization angle $\phi=0$. 
However, it turns out in general that neither $\mathbf{S}$ nor 
$\mathbf{V}$ depend on the angle of polarization $\phi$ (if its value is the same for all
four waves). 
This is understandable because despite the fact that the polarization vectors of the 
quartet do depend on $\phi$ according to Eq.~(\ref{4e}), the combination of 
their mutual relative directions, which determines $\mathbf{S}$ and $w$, is 
invariant.

\section{4-wave fields without regions of negative energy velocity}

A closer study of the effect of the angle $\phi$ on the energy flow reveals that
what is essential is the difference between the angle $\phi_{13}$ for the pair
$(1,3)$ and the angle $\phi_{24}$ for the pair $(2,4)$ in the quartet of
waves. Thus, if $\phi_{13}-\phi_{24}=0$, we return to the results of the
previous section. Of special interest are the cases $\phi_{13}-\phi_{24}=\pm\pi
/2$. Since the polarization vectors together with fields $\mathbf{E}$ and
$\mathbf{B}$ depend on both $\phi_{13}$ and $\phi_{24}$, we prescribe values
to them separately and for simplicity make one of them equal to zero. So,
first we consider the case $\phi_{13}=\pi/2$ and $\phi_{24}=0$. Then
$\mathbf{e}_{1}=\mathbf{e}_{3}=\left(  0,1,0\right)  $ and $\mathbf{e}%
_{2}=\mathbf{e}_{4}=\left(  1,0,0\right)  ,$ i.e., the pair $(1,3)$ is nothing
but a replica of the pair $(2,4)$ rotated about the axis $z $ clockwise by
$\pi/2$. Analogously to Eq.~(\ref{Enurgad0}) and introducing the dimensionless
arguments, we obtain%

\begin{equation}
\mathbf{E}\left(  \widetilde{x},\widetilde{y},\widetilde{z},\widetilde{t}%
\right)  =\left(
\begin{array}
[c]{c}%
2\cos\left(  \widetilde{z}-\widetilde{t}\right)  \cos\widetilde{y}\\
2\cos\left(  \widetilde{z}-\widetilde{t}\right)  \cos\widetilde{x}\\
0
\end{array}
\right) \label{Enurgad90}%
\end{equation}
and likewise for the magnetic field%

\begin{equation}
\mathbf{B}\left(  \widetilde{x},\widetilde{y},\widetilde{z},\widetilde{t}%
\right)  =\left(
\begin{array}
[c]{c}%
-2\cos\left(  \widetilde{z}-\widetilde{t}\right)  \cos\widetilde{x}\cos
\theta\\
2\cos\left(  \widetilde{z}-\widetilde{t}\right)  \cos\widetilde{y}\cos\theta\\
-2\sin\left(  \widetilde{z}-\widetilde{t}\right)  \left(  \sin\widetilde{x}%
-\sin\widetilde{y}\right)  \sin\theta
\end{array}
\right)  .\label{Bnurgad90}%
\end{equation}
We see that in the given case the EM field as a whole is transverse electric
(TE). Eq.~(\ref{Sw}) gives for the Poynting vector
and the energy density the expressions

\begin{equation}
\mathbf{S}\left(  \widetilde{x},\widetilde{y},\widetilde{z},\widetilde{t}%
\right)  =\left(
\begin{array}
[c]{c}%
-4\cos\left(  \widetilde{z}-\widetilde{t}\right)  \sin\left(  \widetilde{z}%
-\widetilde{t}\right)  \sin\theta\cos\widetilde{x}~\left(  \sin\widetilde{x}%
-\sin\widetilde{y}\right) \\
4\cos\left(  \widetilde{z}-\widetilde{t}\right)  \sin\left(  \widetilde{z}%
-\widetilde{t}\right)  \sin\theta\cos\widetilde{y}~\left(  \sin\widetilde{x}%
-\sin\widetilde{y}\right) \\
4\cos^{2}\left(  \widetilde{z}-\widetilde{t}\right)  \cos\theta\left(
\cos^{2}\widetilde{x}+\cos^{2}\widetilde{y}\right)
\end{array}
\right) \label{Snurgad90}%
\end{equation}
and
\begin{eqnarray}
w\left( \widetilde{x},\widetilde{y},\widetilde{z},\widetilde{t}\right) 
&=&2\cos ^{2}\left( \widetilde{z}-\widetilde{t}\right) \left( 1+\cos
^{2}\theta \right) \left( \cos ^{2}\widetilde{x}+\cos ^{2}\widetilde{y}%
\right) +  \nonumber \\
&&2\sin ^{2}\left( \widetilde{z}-\widetilde{t}\right) \sin ^{2}\theta \left(
\sin ^{2}\widetilde{x}-\sin ^{2}\widetilde{y}\right) .  \nonumber
\end{eqnarray}

Conclusions from Eq.~(\ref{Snurgad90}) are the same as those derived from
Eq.~(\ref{Snurgad0}) in the previous section, but with one essential
exception: the 6th conclusion now is that the longitudinal component $S_{z}$
\textit{cannot} take negative values, i.e., the energy flows only forward (of
course, it is assumed that $|\theta|\leq\pi/2$ since otherwise the constituent
plane waves would propagate backward). These conclusions, including the absence of
energy backward flow are confirmed by numerically calculated plots, see figure 4.%

\begin{figure}[h]%
\centering
\includegraphics[width=15cm] {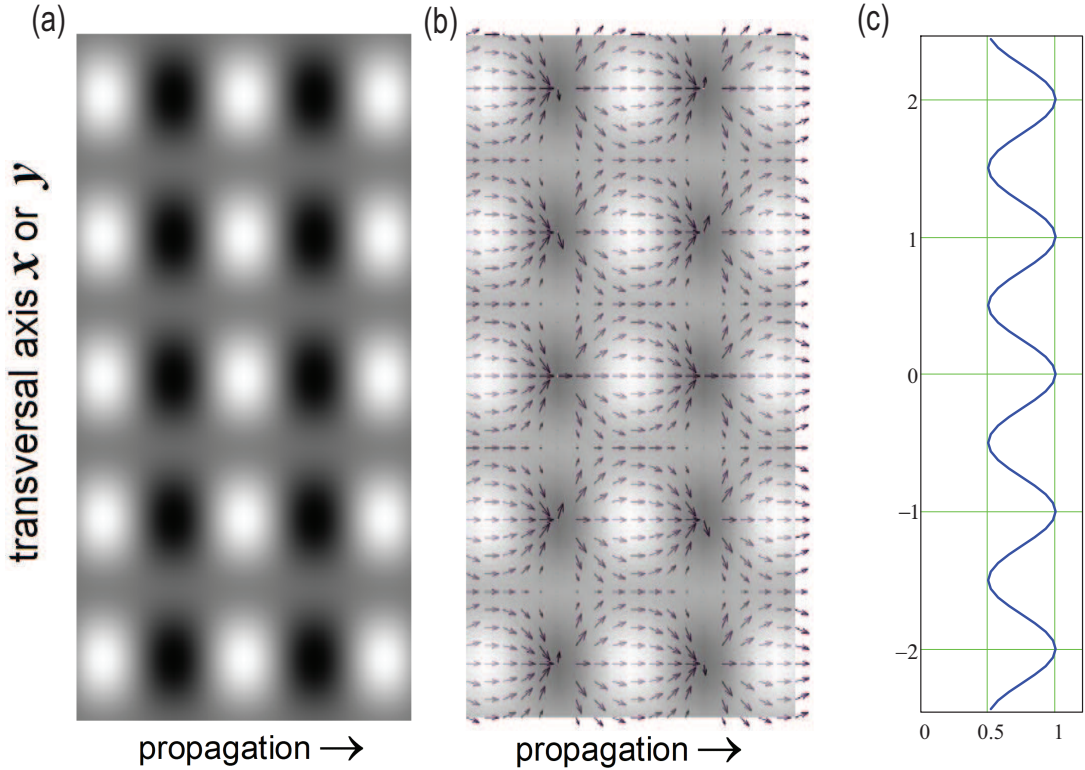}
\caption{Energy density (a) and energy flow (b,c) of 
the quartet of EM plane waves with polarization vectors $\mathbf{e}_{1}%
=\mathbf{e}_{3}=\left(  0,1,0\right)  $ and $\mathbf{e}_{2}=\mathbf{e}%
_{4}=\left(  1,0,0\right)  $ in which case energy flows only in the positive
direction of the axis $z$. For other parameters and scales see the caption 
of figure 2. }%
\end{figure}

As seen in figure 4, all maxima of the energy density are of equal intensity,
the energy velocity vector field is the same at all maxima and contains no
backward flow. But what is very noteworthy is that everywhere the energy
velocity is essentially subluminal (less than $c$) and wherever the energy
flows along the propagation axis $z,$ its velocity is given by the expression%

\begin{equation}
\frac{\left\vert \mathbf{V}\right\vert }{c}=\frac{V_{z}}{c}=\frac{2\beta
}{\beta^{2}+1}=\frac{2\beta^{-1}}{\beta^{-2}+1}~,\label{PSvalem}%
\end{equation}
where $\beta=\cos\theta$. In figure 4, we obtain for $\theta=2\pi/5$ the value $V_{z}=0.564c$.
Precisely the same value comes out from Eq.~(\ref{PSvalem}). For comparison,
as mentioned in the previous section, all fields $\mathbf{E,B,S}$, and $w$
propagate with velocity $v=c/\cos\theta=\beta^{-1}c=3.24c,$ i.e.,
superluminally. Eq.~(\ref{PSvalem}) was first found in \cite{PSvalem} where 
it was shown that various fields---crossing plane waves, Bessel beams,
Bessel-X pulses and other propagation-invariant wave fields obey this relation
between the axial phase- or group velocity and energy flow velocity.

Next, we consider the case $\phi_{13}=0$ and $\phi_{24}=\pi/2$. Then the
polarization vectors in the quartet turn out to be%
\begin{eqnarray}
\mathbf{e}_{1}  & =\left(  \cos\theta,0,-\sin\theta\right)  ,\nonumber\\
\mathbf{e}_{2}  & =\left(  0,\cos\theta,-\sin\theta\right)  ,\nonumber\\
\mathbf{e}_{3}  & =\left(  \cos\theta,0,\sin\theta\right)  ,\label{4eSublum}%
\\
\mathbf{e}_{4}  & =\left(  0,\cos\theta,\sin\theta\right) ,\nonumber
\end{eqnarray}
i.e., the pair $(2,4)$ is nothing but a replica of the pair $(1,3)$ rotated
about the axis $z$ clockwise by $\pi/2$. In contradistinction to the previous
case, the EM field now is transverse magnetic (TM) and the magnetic field turns out
to be equal to Eq.~(\ref{Enurgad90}) but with minus sign in front of the first component.
The expression for the Poynting vector turns out to be the same as in Eq.
(\ref{Snurgad90}) if in the latter the difference of sines is replaced by
their sum and the factor $4$ in the second component by $-4$. As a result, the
quartet with polarization vectors of Eq.~(\ref{4eSublum}) exhibits the same
subluminal velocity of the energy flow as in the previous case and figure 4
illustrates equally well the present case. In \cite{PSvalem} we have shown
that (i) a TM field of a symmetrical pair of plane waves---the
propagating direction of the first one lies on the $(x,z)$ plane and is
inclined by angle $+\theta$ with respect to the $z$-axis, and the second one
by angle $-\theta$ on the same plane---exhibit subluminal energy velocity along
the axis $z$ according to Eq.~(\ref{PSvalem}); (ii) any superpositions of such
pairs rotated about the axis $z$ have the same property.. The quartet of the 
present case is just a particular example of such superposition where the two 
pairs are rotated clockwise by $\pi/2$.

To conclude, from the results of sections 2 and 3 a general rule follows: if
the EM field of the wave quartet is purely TE or TM, there is no backward
energy flow and its (forward) velocity is subluminal obeying
Eq.~(\ref{Snurgad0}); otherwise not only forward but also backward energy flow
takes place, both with \textit{luminal} velocity $c$.

\section{Dependence of energy velocity on polarization angles: general case}

Since the energy flow velocity field so drastically depends on whether $\phi
_{13}=\pi /2$ or $\phi _{13}=0$, it is of interest to study the dependence
over the full range of angle $\phi _{13}=\phi \in \lbrack 0,\pi ]$ (keeping $%
\phi _{24}\ =0)$. Let us focus, in particular, on regions where $\mathbf{V}$
is parallel or antiparallel to the axis $z$ and has magnitude 
$\left\vert \mathbf{V}\right\vert/c=1$ or $V_{z}/c=\pm 1$, see figure 2. 
For the sake of definiteness we consider two points on the axis $x$: 
first, the point where $\widetilde{x}=\widetilde{y}=%
\widetilde{z}-\widetilde{t}=0$ and, second, where $\widetilde{x}=\pi $, $%
\widetilde{y}=\widetilde{z}-\widetilde{t}=0$. For the first point we obtain

\begin{equation}
\frac{V_{z}}{c}=\frac{2\cos \theta +\cos \phi +\cos \phi \cos ^{2}\theta }{%
\cos ^{2}\theta +1+2\cos \phi \cos \theta },  \label{Vfiist}
\end{equation}%
which for $\phi =0$ reduces to $V_{z}/c=1$ as it has to, see figure 2 and
for $\phi =\pi /2$ reduces to Eq.~(\ref{PSvalem}) as it also has to, see
figure 4. For the second point a slightly modified version of Eq.~(\ref%
{Vfiist}) comes out: only a replacement  $\cos \phi \rightarrow -\cos \phi $
must be done. Again, the modified expression for $\phi =0$ reduces to $%
V_{z}/c=-1$ and for $\phi =\pi /2$ to Eq.~(\ref{PSvalem}), as expected.
The behaviour of $V_{z}/c$ over the full range of angle is depicted in figure 5.%

\begin{figure}[h]%
\centering
\includegraphics[width=8cm] {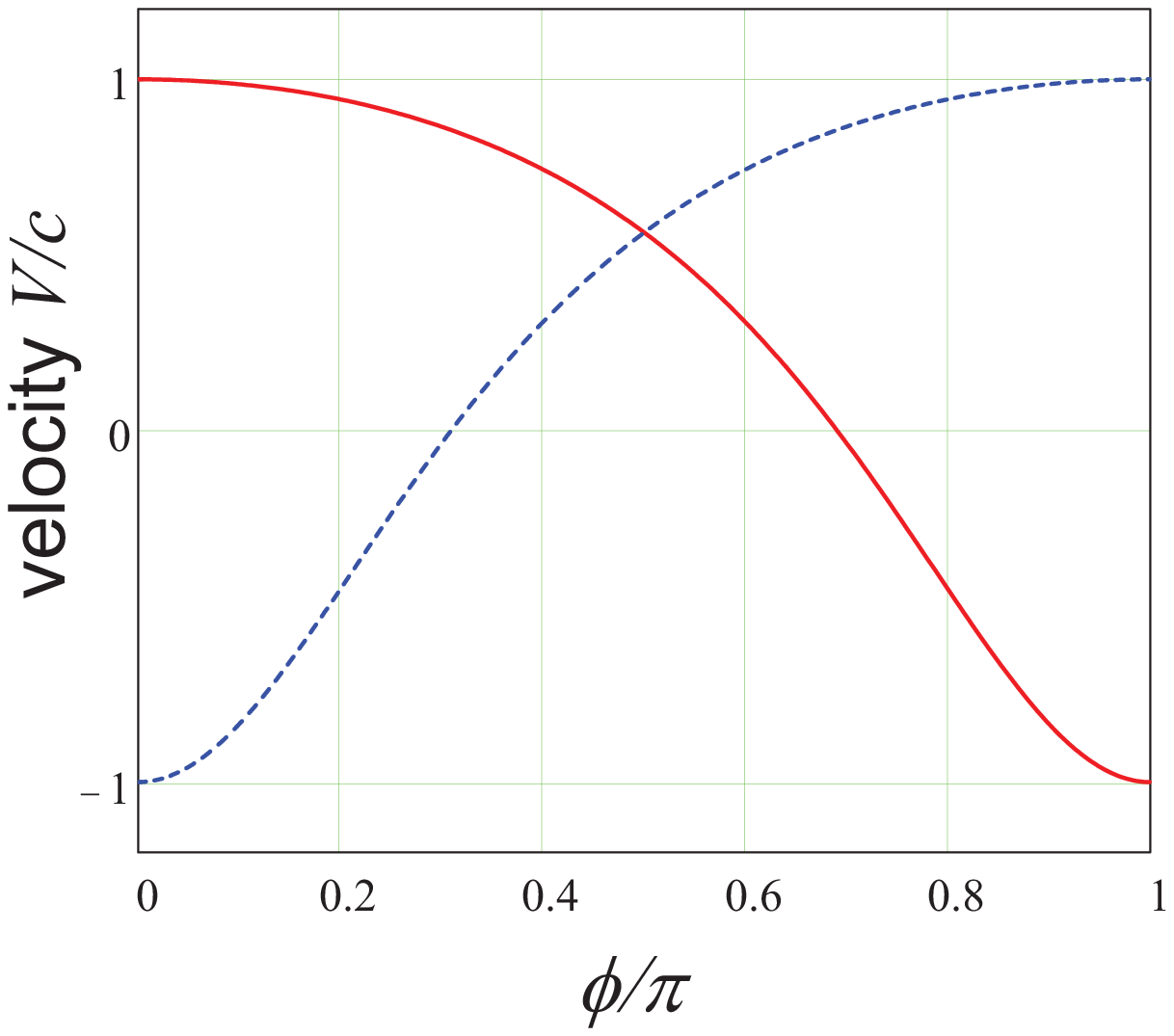}
\caption{Dependence of the energy
flow velocity on the angle $\protect\phi $ for points $\protect\widetilde{x}%
=0$ (solid curve) and $\protect\widetilde{x}=\protect\pi $ (dashed curve)
according to Eq.~(\protect\ref{Vfiist}).}
\end{figure}

We see that if $\phi =\pi /2$ the velocity is given by Eq.~(\ref{PSvalem})
in accord with figure 4. At values $\phi $ close to $\pi $ the regions of
forward and backward flow of energy interchange. Equating the nominator of
Eq.~(\ref{Vfiist}) to zero allows one to answer
the question: what are the values of $\phi $ at which the energy backflow
disappears, i.e., both  curves in figure 5 become positive? The answer is: $%
\phi $ must obey the condition $\cos \phi =\pm 2\cos \theta /\left( \cos
^{2}\theta +1\right) $, i.e., the right-hand-side of Eq.~(\ref{PSvalem}). 
Whether such a surprising coincidence has a deeper physical meaning, needs
further study.

Similarly to the preceding section we also studied the behaviour of the
velocity if the values of angles $\phi _{13}$ and $\phi _{24}$ are
interchanged, i.e. the velocity dependence on $\phi _{24}$ over the full
range of angle $\phi _{24}=\phi \in \lbrack 0,\pi ]$ (keeping $\phi _{13}\
=0)$. Although the interchange of the angles alters the fields $\mathbf{E}$
and $\mathbf{B}$ (interchange their components and signs), the Poynting
vector $\mathbf{S}$, as well as the energy density $w$ and therefore also $%
\mathbf{V}$ remain unchanged. Hence, the results given above hold
in this case as well.

\section{Discussion}

Since the variable along the horizontal axis in figures 2a, 2b, 4a, and 4b is
$z\cos\theta-ct$, these plots can be viewed either as $z$-dependence or time
dependence of the fields. The latter viewpoint allows one to read out from these
plots what would be the temporally cycle-averaged quantities commonly evaluated
from complex-valued EM fields. We see that the direction of energy flow does
not flip in time (and/or along the axis $z$) and, in particular, where it flows
backward in figure 2b, it does so all time. Also, the transverse
components of the velocity vanish as a result of such averaging. This follows
already from symmetry considerations and from the observation that the product
$\cos\left(  \widetilde{z}-\widetilde{t}\right)  \sin\left(  \widetilde{z}%
-\widetilde{t}\right)  $ entering the transverse components in
Eqs.~(\ref{Snurgad0}) and (\ref{Snurgad90}) vanishes after cycle averaging. In
Ref.~\cite{quartet} the cycle-averaged $z$-component $P_{z}$ of the Poynting
vector has been evaluated for the same polarization vectors as in our
Eq.~(\ref{4e}) and plotted on the plane $(z=0)$. Our more detailed results are
in accord with corresponding results in Ref.~\cite{quartet}. An example of
a wave quartet which does not reveal negative $P_{z}$ has been also studied by
the authors of the cited article, but they did it with another geometry of the
polarization vectors than in our two cases of section 3. In their quartet one
pair of the polarization vectors is also a replica of the other rotated about
the axis $z$ clockwise by $\pi/2$; the field is also TM and the cycle-averaged
$z$-component of the Poynting vector is positive everywhere. We studied the 
cycle non-averaged vector field of energy velocity in such quartet and found 
that in contradistinction to figure 4b, the velocity vanishes at the maxima of 
the energy density $w$, whereas its longitudinal component at the minima of $w$ 
is given by Eq.~(\ref{PSvalem}).

The general conclusion from the preceding section is in accord with what has 
been found for the Bessel beams and X-waves \cite{TM+TE,TM+TExwave}: without
superposition of TE and TM fields, there is no reversed Poynting vector. This
is understandable, since the quartet is an elementary example of
superpositions that construct these much more complicated waves.

Although we considered here only monochromatic waves, the results obtained
hold also for ultrashort propagation-invariant pulses, more exactly---for apex
regions of their X-like or double conical field profile. This can be readily
proved by applying the approach used in \cite{PSvalem, meieReactive}]
for the evaluation of the Poynting vector of few-cycle and single-cycle light
sheets. Keeping in mind that the velocity $v=c/\cos\theta$ for such pulses is
simultaneously phase and group velocity, we encounter a striking conclusion:
according to Eq.~(\ref{PSvalem}) the energy flows subluminally while the
pulse itself propagates superluminally! Definitely the statement
\textquotedblleft if an energy density is associated with the magnitude of the
wave... the transport of energy occurs with the group velocity, since that is
the rate of which the pulse travels along\textquotedblright\ (Ref.
\cite{JackKiirus}, sec. 7.8) cannot hold if the group velocity exceeds $c$.
A possibility to resolve this contradiction is offered by the notion of
reactive energy \cite{meieReactive}.

It is well known that the Poynting vector is not defined uniquely by the
Poynting theorem. Could it be that, consequently, the energy-flow velocity we
have dealt with throughout this paper is also not defined uniquely and this is
the reason of the energy backflow effect and non-equality of its velocity to
the group velocity? The discussion of whether the Poynting vector gives the
local energy flow or not has a century-long history, see, e.g., a review
\cite{PoyReview}. In particular, an alternative definition of the Poynting vector 
has been proposed which is more in harmony with the notion of group velocity
\cite{alternPoy}. However, we are not going to dig into the problem here and
simply refer to \cite{JackKiirus} (sec. 6.7), where it is stated that if one 
takes into account that the Poynting vector defines also the momentum density of 
the field and the definition must not violate the theory of relativity,
the common expression for the Poynting vector is unique.

An intriguing alternative definition of the energy velocity is proposed in
\cite{Johns2020}. It springs forth from an idea that a correct velocity in a given
location must be equal to the velocity of an inertial frame in which energy
flux turns to be zero at the given location. Such approach leads to the
following relation between the magnitude $V\equiv\left\vert \mathbf{V}%
\right\vert =\left\vert \mathbf{S}\right\vert /w$ of the commonly defined
energy velocity and its alternative version $V_{a}$%
\begin{equation}
\frac{V}{c}=\frac{2\left(  V_{a}/c\right)  }{1+\left(  V_{a}/c\right)  ^{2}%
}~.\label{John11}%
\end{equation}
This relation, which is Eq.~(11) from Ref.~\cite{Johns2020} rewritten in our
designations, surprisingly coincides with Eq.~(\ref{PSvalem}) if we equate
$V_{a}/c$ to the normalized group velocity $\beta^{-1}$ or to its reciprocal.
Indeed, in the reference frame moving with $V_{a}=\beta$ along the axis $z$
the angle $\theta$ transforms to $\pi/2$ not only for the wave quartet but also
for Bessel beams and X-type pulsed localized waves \cite{meie2xLor}
resulting according to Eq. (\ref{Snurgad90}) in vanishing of $S_{z}$. However,
further analysis of the alternative definition of the energy flow velocity is
outside the scope of this study.

\section{Conclusion}
The results of this study of the electromagnetic field of four interfering plane waves 
demonstrate in detail how crucial is the polarization of the waves for the existence of regions where the energy flows backward with respect to the propagation direction of 
the field itself. Moreover, by adjusting the relative angle between linear polarizations of two pairs in the quartet of waves, one can obtain any value between $-c$ and $c$ of the energy flow velocity at maxima or minima of energy density.

Generalizations of the results for interfering few-cycle ultrashort pulses is straightforward. They may also be useful in studies of quantum backflow and optical analogues of this intriguing effect in particle physics \cite{Q1}-\cite{Q3}.

\end{document}